\documentclass[aps,prl,twocolumn,reprint,preprintnumbers,floatfix,nofootinbib]{revtex4}

\usepackage{amsmath}
\usepackage{amssymb}

\usepackage{graphicx}
\usepackage{dcolumn}
\usepackage{bm}
\usepackage{marvosym}
\usepackage{tabulary}

\usepackage{color}
\definecolor{light-gray}{gray}{0.5}
\definecolor{blue}{rgb}{0.0,0.0,1.0}
\definecolor{green}{rgb}{0.0,0.5,0.0}
\definecolor{red}{rgb}{1.0,0.0,0.0}
\definecolor{cyan}{rgb}{0.0,0.75,0.75}
\definecolor{magenta}{rgb}{0.75,0.0,0.75}
\definecolor{yellow}{rgb}{0.75,0.75,0.0}

\newcommand{\avg}[1]{\langle{#1}\rangle}

\newcommand{\lrbig}[1]{\left( {#1} \right)}

\newcommand{\dd}{\mathrm{d}}

\begin{document}
\title{Rapid growth of cloud droplets by turbulence}
%
\author{V. Dallas}
\author{J. C. Vassilicos}
\affiliation{Department of Aeronautics, Imperial College, London, SW7 2AZ, UK}
%
%
\begin{abstract}
Abstract: Assuming perfect collision efficiency, we demonstrate that
turbulence can initiate and sustain rapid growth of very small water
droplets in air even when these droplets are too small to cluster, and
even without having to take gravity and small-scale intermittency into
account. This is because the range of local Stokes numbers of
identical droplets in the turbulent flow field is broad enough even
when small-scale intermittency is neglected. This demonstration is
given for turbulence which is one order of magnitude less intense than
typically in warm clouds but with a volume fraction which, even though
small, is nevertheless
large enough for an estimated a priori frequency of collisions to be
ten times larger than in warm clouds. However, the time of growth in
these conditions turns out to be one order of magnitude smaller than
in warm clouds.


\end{abstract}
\maketitle
%

The rapid growth of warm (ice-free) cloud droplets from $15 \mu m$ to
about $50 \mu m$ in a short time, typically half an hour, is a
well-known phenomenon which still defies explanation. This phenomenon
is essential for understanding rain initiation and cloud albedo
\cite{shaw03, baker97}. 

Condensation dominates cloud droplet growth till about 10 to 15 $\mu
m$ and may be producing a narrow droplet size distribution
\cite{wangetal06, shaw03, falkovichpumir07}. If so, a subsequent growth mechanism
involving similar-sized particles is required to make the droplet size
distribution wide enough for different-sized gravitational collisions
to be effective in a final droplet growth mechanism. Such a
gravitational settling mechanism may dominate from 30 to 50
micrometres and above \cite{wangetal06}. The identification of what triggers the
intermediate growth between $15 \mu m$ to about $50 \mu m$ and the
rate with which it proceeds are major challenges of cloud physics.

Various authors have already proposed that turbulence may be the basis
of the intermediate mechanism whereby initially monodispersed droplets
can grow via turbulence-generated collisions \cite{wangetal00, falkovichpumir07}. This mechanism is the specific concern of the present
paper. It has been suggested that turbulence in clouds may be
generating preferential concentrations (clustering) of droplets which
would cause a sharp increase in collision and coalescence events and
therefore a fast growth of droplet sizes \cite{shaw03, sundaramcollins97, zhouetal98b}.
It has also been suggested that caustics may activate such fast
droplet growth \cite{wilkinsonetal06}. However, measurements suggest
\cite{vaillancourtyau00} that, in warm clouds, the droplet response
time $\tau_p$ is much too small compared to the smallest (Kolmogorov)
time scale $\tau_{\eta}$ for any significant preferential
concentration or caustics to be observable and meaningful. Indeed, 
Saffman \& Turner \cite{saffmanturner56} considered the case
where the Stokes number $\tau_{p}/\tau_{\eta}$ is effectively zero but
the droplet size is finite and much smaller than the Kolmogorov
micro-length-scale $\eta$ (the smallest length scale of the
turbulence). They showed that, in this case, droplet sizes do not grow
fast enough to explain cloud dynamics and statistics. Their assumption
on the droplet size is accurate as $\eta \sim 1mm$ and droplet radius
$a_{p} \sim 10^{-2}mm$ in clouds before rain initiation and it is
reasonable to assume, as they did, that such very small droplets are
spherical. However, they also assumed that turbulent velocity
gradients are statistically gaussian, and this is known not to be
true. Small-scale turbulence is intermittent and the turbulent
velocity gradients are increasingly non-gaussian as the Reynolds
number increases \cite{pope00}.

The neglect of intermittency may be a significant shortcoming because
rare but powerful turbulence events could cause neighbouring droplets
to collide and coalesce faster than one would expect from a
consideration of the average properties of the turbulence field. These
collisions could generate a few large droplets with high momentum
which could trigger a chain of successive collisions when travelling
and falling fast through the field of much smaller droplets. In
principle, such a chain reaction could lead to significant droplet
size growth. Kostinski \& Shaw \cite{kostinskishaw05} have
already argued that rare but powerful events are required for droplet
growth, and that these events may have their cause in small-scale
turbulence intermittency. Ghosh et al. \cite{ghoshetal05} have
argued that such rare but powerful small-scale eddies can also
selectively increase settling velocities and thereby further enhance
droplet size growth rates.


In this paper we show that turbulence can generate fast droplet size
growth without the need for small-scale intermittency and differential
gravitational settling, even when droplets are too small to
cluster. We place ourselves in a situation close to but different from
Saffman \& Turner's \cite{saffmanturner56}. Close in the sense that we assume gaussian
statistics of turbulent velocity gradients, spherical droplets of
finite size much smaller than $\eta$ and a particle response time
$\tau_p$ which, as a result, is very small. However our analysis
differs from that of Saffman \& Turner \cite{saffmanturner56}
in that it takes into account the broad range of local micro-time
scales of the turbulence and therefore a broad range of local Stokes
numbers in the flow. Nevertheless, all these Stokes numbers are
predominantly too small for any reasonable level of clustering to be
clearly present.

We follow the approach of Reade \& Collins \cite{readecollins00a} 
who used a Direct Numerical Simulation (DNS) of
the incompressible Navier-Stokes equations to generate the turbulent
velocity field in which they integrated trajectories of very small
spherical particles with high mass density. They then applied the
algorithm of Sundaram \& Collins \cite{sundaramcollins96} to
simulate droplet collisions and coalescence. However, their DNS was of
3D homogeneous isotropic turbulence without well-defined inertial
range ($Re_\lambda = 55$) and the initial Stokes numbers of their
droplets ranged between 0.2 and 0.7 which is large enough for
clustering to occur (see Fig. 1 in \cite{readecollins00b} and Fig. 1 in \cite{cgv06}). 
Instead, our Stokes numbers are initially all well
below 0.1 and we opt for a DNS of 2D inverse-cascading turbulence
which ensures a wide $-5/3$ energy spectrum, much wider than can be
achieved by 3D DNS. Also, it is well known that the velocity gradient
statistics of 2D homogeneous isotropic turbulence are approximately
gaussian \cite{tabeling02}. In other words, there is no small-scale
intermittency in the turbulence we are using.

\begin{figure}[!h]
  \centering 
  \includegraphics[width=8.5cm]{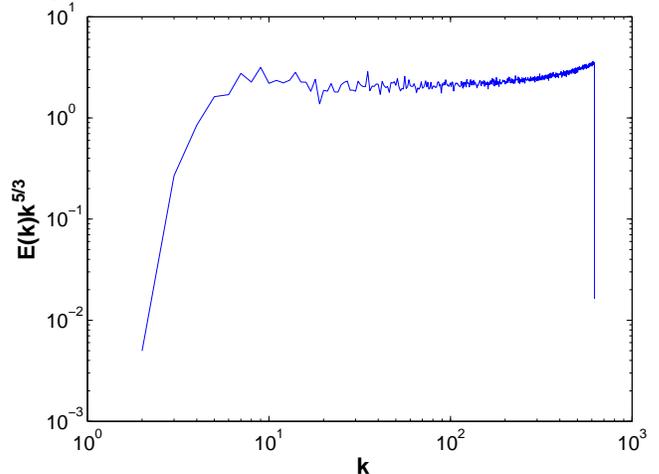}
  \caption{(Color online) Energy spectrum compensated with $k^{5/3}$ and obtained
    from a $2048^2$ inverse-cascading 2D turbulence using the
    numerical method and high wavenumber filtering of \cite{gv04}. The
    low-pass filtering operation and its wavenumber $k_c$ are clearly
    visible on the right side of the plot. This operation removes the
    scales directly affected by the small-scale forcing needed to give
    rise to a stationary turbulence with a $k^{-5/3}$ energy spectrum
    at lower wavenumbers $k$.}
  \label{fig:spectrum}
\end{figure}

We endeavour to use a set of conditions as close to real warm clouds
as possible within the extent allowed by our computational
capabilities. The ratio of the outer to the inner length-scales of our
2D turbulence is at least $L/\eta = 110.8$. This is way below the
ratio $10^5$ in warm clouds, but high enough to have a wide inertial
range (see Fig. \ref{fig:spectrum}). The rms turbulence velocity $u'$
is set at a value which ensures that if our numerical values of $\eta$
and $\tau_{\eta}$ are taken to correspond to $1\,mm$ and $10^{-1} -
10^{-2}\,s$ respectively as in warm clouds, then $u'$ is smaller than
the usual value in warm clouds which hovers around $1\,m/s$. In fact,
our value of $u'$ is an order of magnitude smaller which makes it
harder for our simulations to produce collisions and also allows the
adoption of small Stokes numbers, much smaller than 0.1 as in warm
clouds, without having to take inordinately small time steps in our
time-integrations.


In warm clouds, the ratio of the droplet's water density to the
density of the surrounding air is $\rho_p / \rho_f \sim 10^3$ and the
ratio of the mean droplet radius to the Kolmogorov micro-length scale
before rain initiation is $\avg{a_p}/\eta \sim 10^{-2}$. In our
simulations we take $\rho_p / \rho_f = 10^3$ and initial
$\avg{a_p}/\eta = 0.0135$ where $2\pi/\eta$ is taken to be a large
multiple of the filter wavenumber $k_c$ (see Fig. \ref{fig:spectrum}). The larger we chose this multiple to be, the
smaller $\avg{a_p}$ becomes and the larger the number of droplet
trajectories needed to be integrated if we want to keep a realistic
droplet volume/area fraction. The droplet volume fraction is $\phi_{V}
\sim 10^{-6}$ in warm clouds but we need to set an equivalent area
fraction $\phi_A$ in our 2D simulations. We do this by requiring that
the geometrical probability of interception along straight lines
between two droplets
is the same in 2D and 3D. Given a number $N_2$ of homogeneously
distributed droplets of radius $a_p$ in an area $L_{box}^2$ in 2D
space, a rough estimate of this 2D probability is $4 \cdot 3 \cdot
2a_p / (2\pi l_2) = 12 a_p / (\pi l_2)$ where $l_2 =
(L_{box}^2 / N_2)^{1/2}$ is the average distance between droplets
(about 4 droplets at a distance $l_2$ from each droplet). Given a
number $N_3$ of homogeneously distributed droplets of radius $a_p$ in
a volume $L_{box}^3$ in 3D space, a rough estimate of this 3D
probability is $6\pi(2a_p)^2 / (4\pi l_3^2) = 6(a_p / l_3)^2$
where it is $l_3 = (L_{box}^3 / N_3)^{1/3}$ which is now the average
distance between droplets (about 6 droplets at a distance $l_3$ from
each droplet). Equating the two probabilities yields $\phi_A =
{\pi \over 4} \lrbig{9\pi \over 16}^{2/3}
\phi_V^{4/3}$. Equivalently, this means that the number of particles
which we need to immitate a certain 3D volume fraction $\phi_V$ in
our 2D simulations is
\begin{equation} 
N_2 = {1 \over 4} \lrbig{9\pi \over 16}^{2/3}
\phi_V^{4/3} \lrbig{L_{box} \over a_p}^2.
\end{equation}
Our parameters are tabulated and compared below with the
parameters of typical clouds.


\begin{widetext}
 \begin{center}
  \begin{table}[!ht]
   \begin{ruledtabular}
    \begin{tabular}{*{13}{c}}
     & $N$  & $L/\eta$ & $u'$ & $\overline{\tau}_\eta$ & $L / (u'\overline{\tau}_\eta)$ & $St$ & $\avg{a_p}/\eta$ & $\phi_V$  & $\phi_A$ & $\rho_p / \rho_f$ \\
      \hline
     Clouds & - & $10^5$ & $1\,m/s$ & $0.1\,s$  & $10^3$ & $10^{-4}-10^{-2}$ & $10^{-2}$ & $10^{-6}$ & - & $10^3$ \\
     DNS & $2048^2$ & $110.8$ & $1.24$ & $0.0074$ & $26.7$ & $0.04$ & $0.0135$ & - & $10^{-4}$ & $10^3$
    \end{tabular}
    \end{ruledtabular}
    \caption{Comparison of parameters in typical warm clouds and in our
  DNS. The particle response time $\tau_{p}= {2\rho_{p} \over
    9\rho_{f}} ({a_p \over \eta})^2 \overline{\tau}_{\eta}$, and we define
  the Stokes number $St = {2\rho_{p} \over 9\rho_{f}}
  ({\avg{a_p} \over \eta})^{2}$, where the brackets are now an average over
  all particles/droplets. (In keeping with $\rho_p$ and $\rho_f$ which
  are mass densities in a volume, these expressions for $\tau_p$ and
  $St$ are for 3D spheres, not 2D disks. It is more important to keep
  a realistic dependence on $a_p$ than 2D consistency in our model.)}
   \label{tbl:parameters}
  \end{table}
 \end{center}
\end{widetext}

\begin{figure}[!h]
  \centering
  \includegraphics[width=8.5cm]{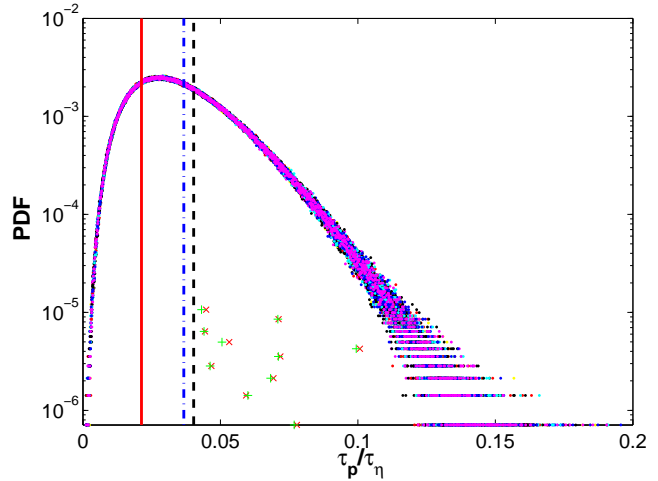}
  \caption{(Color online) PDFs of $\tau_p /\tau_\eta$ at $t = 0$ and at the times of
    the first 10 collisions for the case where $\phi_A = 10^{-4}$
    and $\eta/(2\pi/k_c) = 1/5$, see Table \ref{tbl:parameters}. The
    red solid, blue dashed-dotted and black dashed lines indicate where $\avg{\tau_p }/
    \avg{\tau_\eta} =2.13 \cdot 10^{-2}$, $\avg{\tau_p /
      \tau_\eta} = 3.67 \cdot 10^{-2}$ and $\avg{\tau_p}/
    \overline{\tau}_\eta = 4.03 \cdot 10^{-2}$ are, respectively,
    on the plot. The pairs of red crosses and green plus signs on the plot
    correspond to early collisions and indicate, on the abscissa, the
    values of the local Stokes numbers $\tau_p /\tau_\eta$ of the
    two colliding particles one time step before collision. The
    ordinate value of these crosses and plus signs is arbitrary except for the order
    of collisions, the red cross and the green plus sign corresponding to the first (tenth)
    collision having the lowest (highest) ordinate value. The fact
    that crosses and plus signs for the same collision are always very
    close to each other reflects the fact that the first ten
    collisions are for similar-size particles.}
\label{fig:Stinterm}
\end{figure}

The initial particle size distribution in our simulations is a narrow
log-normal with a mean particle radius $\avg{a_p} = 0.0135\eta$ and a
width $\delta a_p $ between smallest and largest size such that
$\avg{a_p} / \delta a_p $ is about 10 or larger (we tried up to 50 and
did not find any differences in our conclusions).  The Kolmogorov time
scale ${\overline \tau}_{\eta}$ is an average time scale determined by
the average turbulent kinetic energy dissipation rate per unit mass.
This average dissipation rate is proportional to $2 \avg{tr ({\bm
    s}^{2})}$ where ${\bm s}$ is the strain rate tensor and the
brackets are an average over all space. Hence, $\overline{\tau}_{\eta}
= 1/\sqrt{2\avg{tr({\bm s}^{2})}}$.  The local micro-time scales
$\tau_{\eta}$ of the turbulence are determined in the same way but in
terms of the local turbulent kinetic energy dissipation rates per unit
mass which are proportional to the local $2 tr ({\bm s}^{2})$. In
other words the relevant local micro-time scales are determined by the
local strain rates such that $\tau_{\eta} = 1/\sqrt{2 tr ({\bm
    s}^{2})}$. The Probability Density Function (PDF) of all Stokes
numbers $\tau_p /\tau_{\eta}$ in our flow is given in Fig.
\ref{fig:Stinterm}. The red solid, blue dashed-dotted and black dashed lines on this plot
mark the values of $\tau_p /\tau_{\eta}$ which equal $\avg{\tau_p} /
\avg{\tau_{\eta}}$, $\avg{\tau_p / \tau_{\eta}}$ and $\avg{\tau_p} /
\overline{\tau}_{\eta}$ respectively (the brackets being averages over
all particles or all space accordingly).
We checked that the PDF of $\tau_p / \tau_\eta$ is very similar to the
PDF of $1 / \tau_\eta$ as expected from the fact that the PDF of
$\tau_p$ is very narrowly peaked, and we also checked that the PDF of
$tr ({\bm s}^2)$ is peaked at zero, which agrees with the observation
that the PDFs of $\tau_p / \tau_\eta$ and of $1 / \tau_\eta$ both
vanish at zero. Finally, we verified that the PDFs of partial
derivatives of velocity components with respect to spatial coordinates
are approximate gaussians peaked at 0.

Following Reade \& Collins \cite{readecollins00a}, droplets are
modelled as very small but dense circular inertial particles
subjected to Stokes drag and impulsive forces $\bm F_p^{ij}$ resulting
from collisions between $i$th and the $j$th particles. These particles
are in fact point particles in the simulations evolving according to
\begin{gather}
 \frac{\dd \bm x_p^i}{\dd t} = \bm v_p^i \\
 \frac{\dd \bm v_p^i}{\dd t} = \frac{1}{\tau_p^i}(\bm u(\bm x_p^i , t) - \bm v_p^i) + \frac{1}{m_p^i} \sum_{j\neq i} \bm F_p^{ij}
\end{gather}
where ${\bm x_p^i}$, $\bm v_p^i$, $m_p^i$ and $\tau_p^i =
\frac{2}{9}\frac{\rho_p}{\rho_f}\frac{(a_p^i)^2}{\nu}={2\rho_p \over
9 \rho_f} ({a_p^i \over \eta})^2 \overline{\tau}_\eta$ are the
positions, velocities, masses and response times of particles of
radius $a_p^i$, and $\bm u(\bm x_p^i , t)$ are the carrier fluid
velocities at ${\bm x_p^i} (t)$ at time $t$.

A method based on molecular-dynamic-simulation strategies
\cite{sundaramcollins96,allentildesley89} checks for collisions
between particles. Collisions are enacted in a random order whenever
two point particles get so close that
 \begin{equation}
  \|\bm x_p^i-\bm x_p^j\| \leq (a_p^i + a_p^j).
  \end{equation}
It is commonly accepted that the coalescence efficiency is close to
one for droplets with radius less than $100 \mu$m because their ratios
of inertial force to surface tension are small
\cite{wangetal08}. However, collision efficiencies may be estimated to
be of the order of a few tens of percent for droplets of $15\mu$m
radius or so \cite{jonas96, pruppacherklett98} even if we take into
account turbulent enhancement of these efficiencies \cite{wangetal08}. 
Nevertheless, as an initial simplifying
assumption, we take collision efficiencies to be 1 in this study.  All
our collisions are therefore assumed to give rise to a coagulation
event thus resulting in a new circular particle with the following
properties
  \begin{gather}
   m_p^{i,new} = m_p^i + m_p^j \\
   \bm v_p^{i,new} = (m_p^i\bm v_p^i + m_p^j\bm v_p^j)/m_p^{i,new}.
   \end{gather}
The positions and velocities of the remaining (non-colliding)
particles are then advanced using $4th$-order Runge-Kutta
algorithm. To ensure accurate integrations of particle trajectories,
the timestep $\Delta t$ used in our DNS satisfies $\Delta t \ll
min_{i} (\tau_{p}^{i} ) \ll \eta/u' < \overline{\tau}_{\eta} <
\avg{\tau_{\eta}}$. This sets a lower bound to our choice of $\eta$
and therefore a lower bound to the volume fraction $\phi_V$ which we
can emulate via equation (1) for a certain number $N_2$ of initial
droplets. Indeed, ${L_{box}\over \avg{a_{p}}} = {L_{box}\over L} {L
  \over 2\pi/k_{c}} {2\pi/k_{c} \over \eta} {\eta \over \avg{a_{p}}}$,
and in our simulations ${L_{box}\over L} = {2\pi\over 0.245} = 25.65$,
${L \over 2\pi/k_{c}} \approx 22.18$ and ${\eta \over \avg{a_{p}}}
\approx 74.1$ for the initial distribution of droplet sizes. For
equation (1) to give us a realistic volume fraction $\phi_{V} \sim
10^{-6}$ would require either extremely small initial $N_{2}$ or
extremely large ${2\pi/k_{c} \over \eta}$ which our time stepping does
not allow. (A very large ${2\pi/k_{c} \over \eta}$ would also require
a very much larger DNS with a very much wider range of excited scales
to be justified.) Furthermore, very small initial values of $N_2$
require much longer integration times than we can afford. We have
therefore opted for the following three sets of simulations: $N_2
\approx 1.5 \cdot 10^6$ and $\eta/(2\pi/k_{c}) = 1/5$ (case of
Table I); $N_2 \approx 5.5 \cdot 10^5$ and $\eta/(2\pi/k_{c}) = 1/10$; $N_2
\approx 1.25 \cdot 10^5$ and $\eta/(2\pi/k_{c}) = 1/15$. In the first case,
equation (1) gives $\phi_V \approx 10^{-3}$, and in the second
and third cases equation (1) gives $\phi_V \approx 2 \cdot 10^{-4}$
and $\phi_V \approx 3 \cdot 10^{-5}$ respectively. Hence, we always
overestimate typical volume fractions in warm clouds by two to three
orders of magnitude, but we also underestimate the turbulence
intensity by a factor of 10. Defining an a priori frequency of
collisions as $u'/L_{box}$ times the geometrical probability of
interception introduced in the text leading to equation (1), this a
priori frequency is $f_{2} \equiv {u'\over L_{box}} {12a_{p}\over \pi
  l_{2}}$ in our simulations and, for the case corresponding to
$\phi_{V} = 10^{-3}$, turns out to be 10 times smaller than in 3D warm
cloud conditions where the volume fraction is $\phi_{V} = 10^{-6}$ and
the turbulence intensity is 10 times larger than here, i.e. $f_{2}
\approx 10 f_{3}^{wc}$.

In the case where $\eta/(2\pi/k_{c}) = 1/5$ and $\phi_{V} = 10^{-3}$,
we integrate the trajectories of about 1.5 million initial particles
for 22 outer time-scales $\tau_{L}\equiv L/u'$ of our flow. Based on
the parameters of Table \ref{tbl:parameters} which are for this case,
and assuming that $\eta$ is about $1\,mm$ and $\overline{\tau}_\eta$
is about $0.1\,s$, this total integration time corresponds to less
than a minute, which is extremely short compared to the typical
fifteen minutes to half hour usually required for droplets to grow
from about $15 \mu$m to about $50 \mu$m in warm clouds. Note that $22
\tau_{L} f_{2}$ is of the same order as fifteen minutes multiplied by
$f_{3}^{wc}$.

\begin{figure}[!h]
  \centering
  \includegraphics[width=0.235\textwidth]{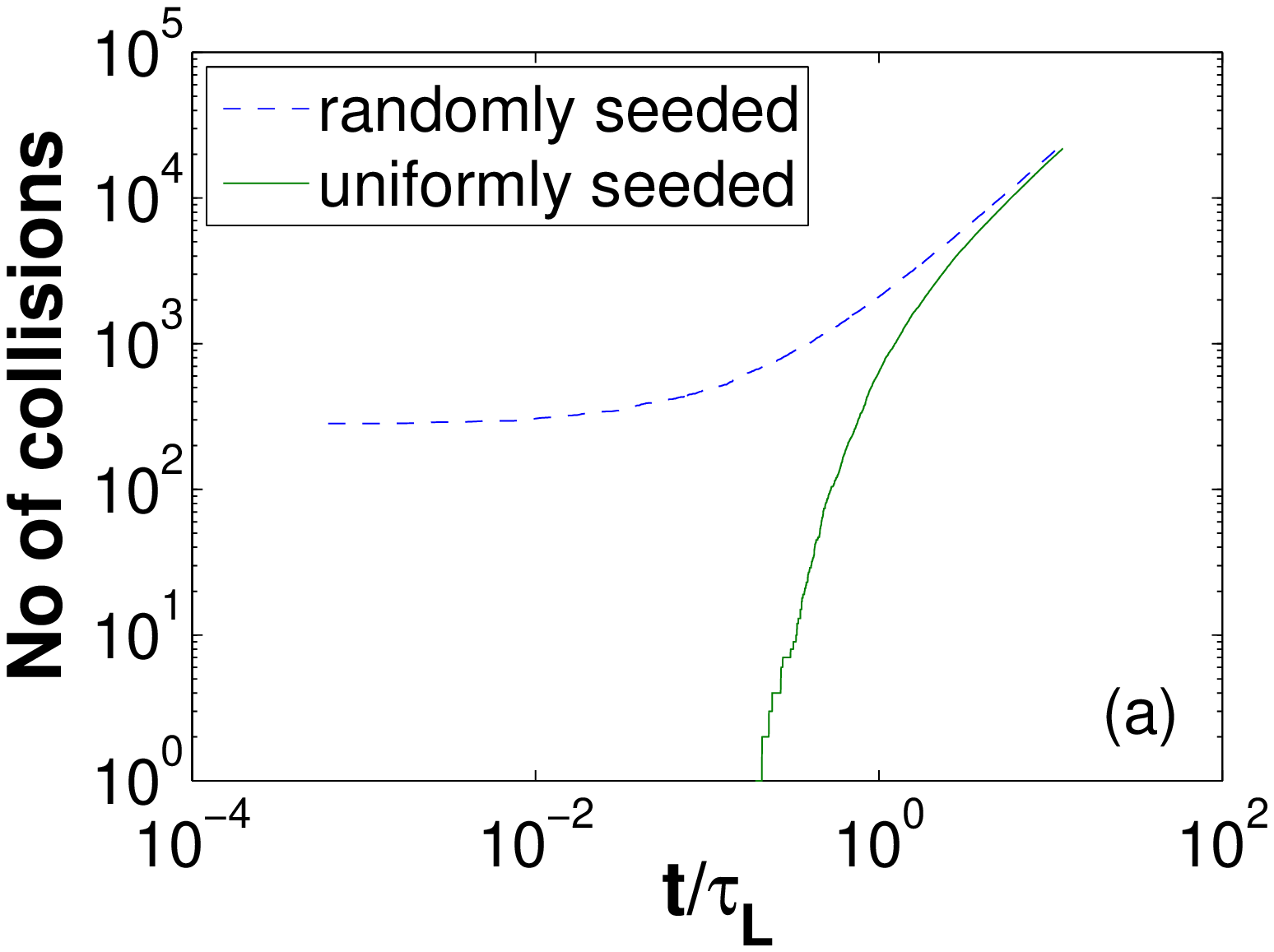}
  \includegraphics[width=0.235\textwidth]{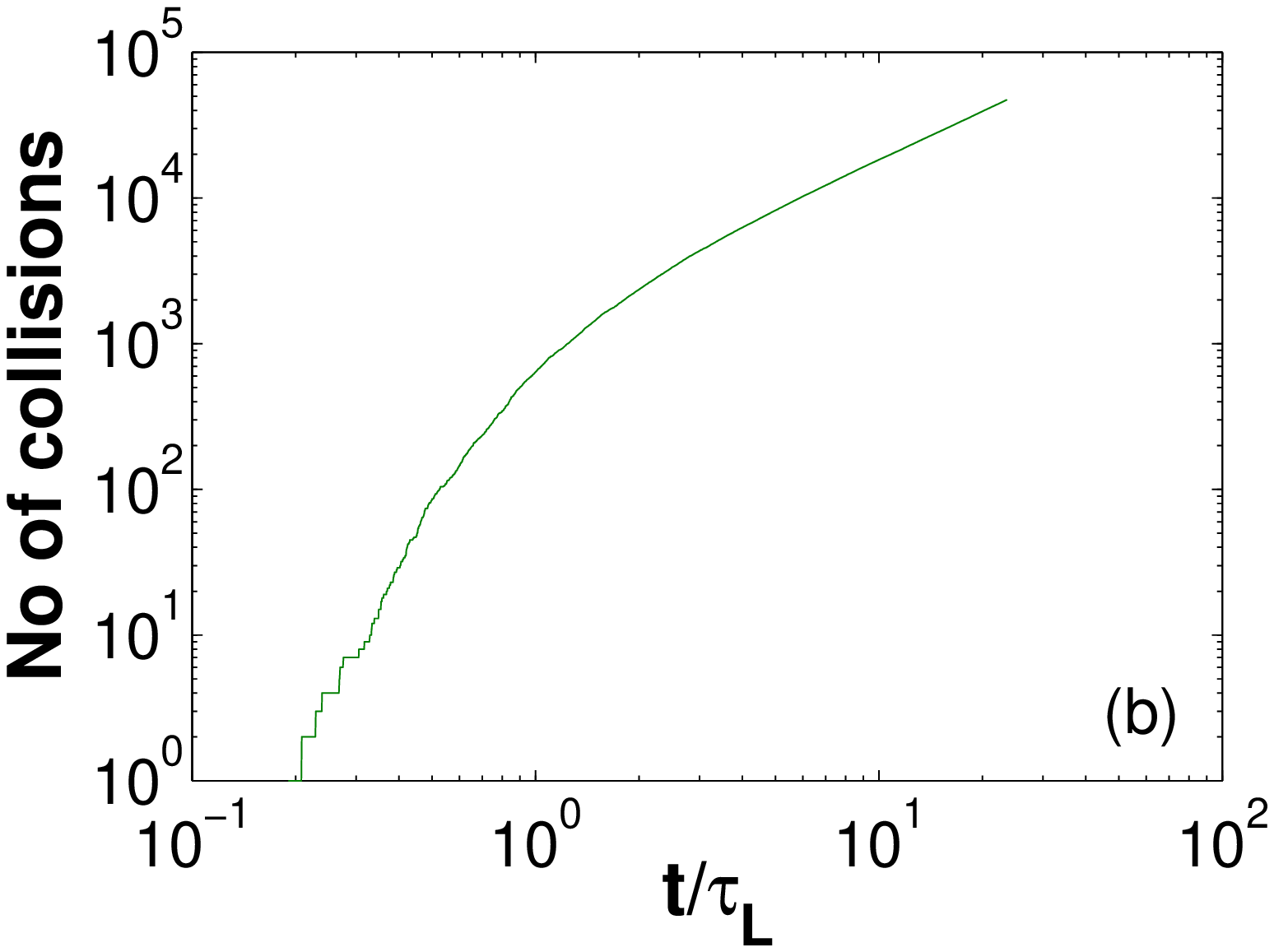}
  \caption{(Color online) Numbers of collisions of pairs of particles as functions of
    $t/\tau_{L} = tu'/L$ for the case where $\phi_{A} = 10^{-4}$ (see
    Table \ref{tbl:parameters}). (a) Evolution up to $t/\tau_{L} = 11$
    for two different initial distributions of particles. (b)
    Evolution up to $t/\tau_{L} = 22$ for the initially uniform
    distribution of particles.}
  \label{fig:collgrowth}
\end{figure}

Numbers of collisions as a function of time are plotted in Fig.
\ref{fig:collgrowth}. We consider two different types of initial
conditions. One where the point particles are randomly distributed, in
which case collisions occur immediately after $t=0$ because a sizeable
number of point particles find themselves, initially, close enough for
(4) to hold. And one where the point particles are distributed on a
regular lattice so that (4) does not hold initially for all pairs of
particles. It might be interesting to note (see Fig.
\ref{fig:collgrowth}) that, whilst the initial evolution of the number
of collisions is very different in these two cases, they converge
towards a similar time dependence at large enough times $t / \tau_L =
t u' / L$.

In the case where particles are uniformly distributed
at $t = 0$, the first collision does not occur immediately, but at a
time equal to about $5.1 \overline{\tau}_\eta \approx 2.67
\avg{\tau_{\eta}}$. This collision is between particles that have a
local Stokes number $\tau_p /\tau_{\eta}$ which is between 2 and 4
times larger than the average Stokes number, depending on the way one
choses to estimate it, see Fig. \ref{fig:Stinterm}. All ten first
collisions occur between time $t=5.1 \overline{\tau}_\eta \approx 2.67
\avg{\tau_{\eta}}$ and time $t=8.8\overline{\tau}_\eta \approx 4.61
\avg{\tau_{\eta}}$ after $t=0$ and involve pairs of particles with Stokes
numbers well above any estimate of the average Stokes number. They
therefore occur within about a second and are caused by extreme events
within the air turbulence where, locally, the Stokes number is higher
than average. Considerations based only on the average Stokes number
would miss these collisions and would therefore also miss the process
initiating droplet size growth. Note, in particular, that the local
Stokes number characterising the first and fifth collisions are
respectively 0.08 and 0.1, just about large enough for meagre signs of
clustering to be present if the average Stokes number had such a value
(see Fig. 1 in \cite{readecollins00b} and Fig. 1 in \cite{cgv06}). 

\begin{figure}[!h]
  \centering
  \includegraphics[width=0.235\textwidth]{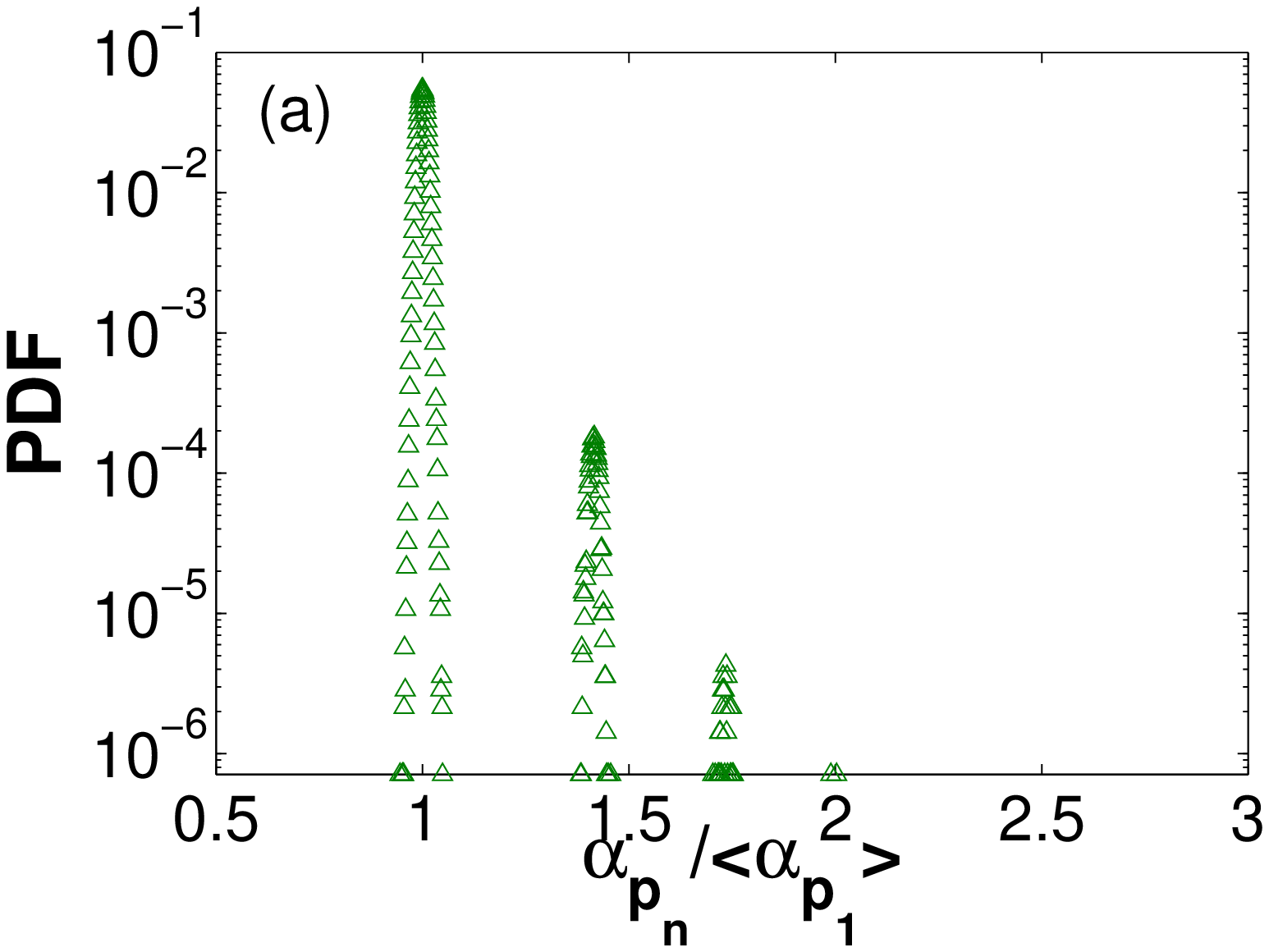}
  \includegraphics[width=0.235\textwidth]{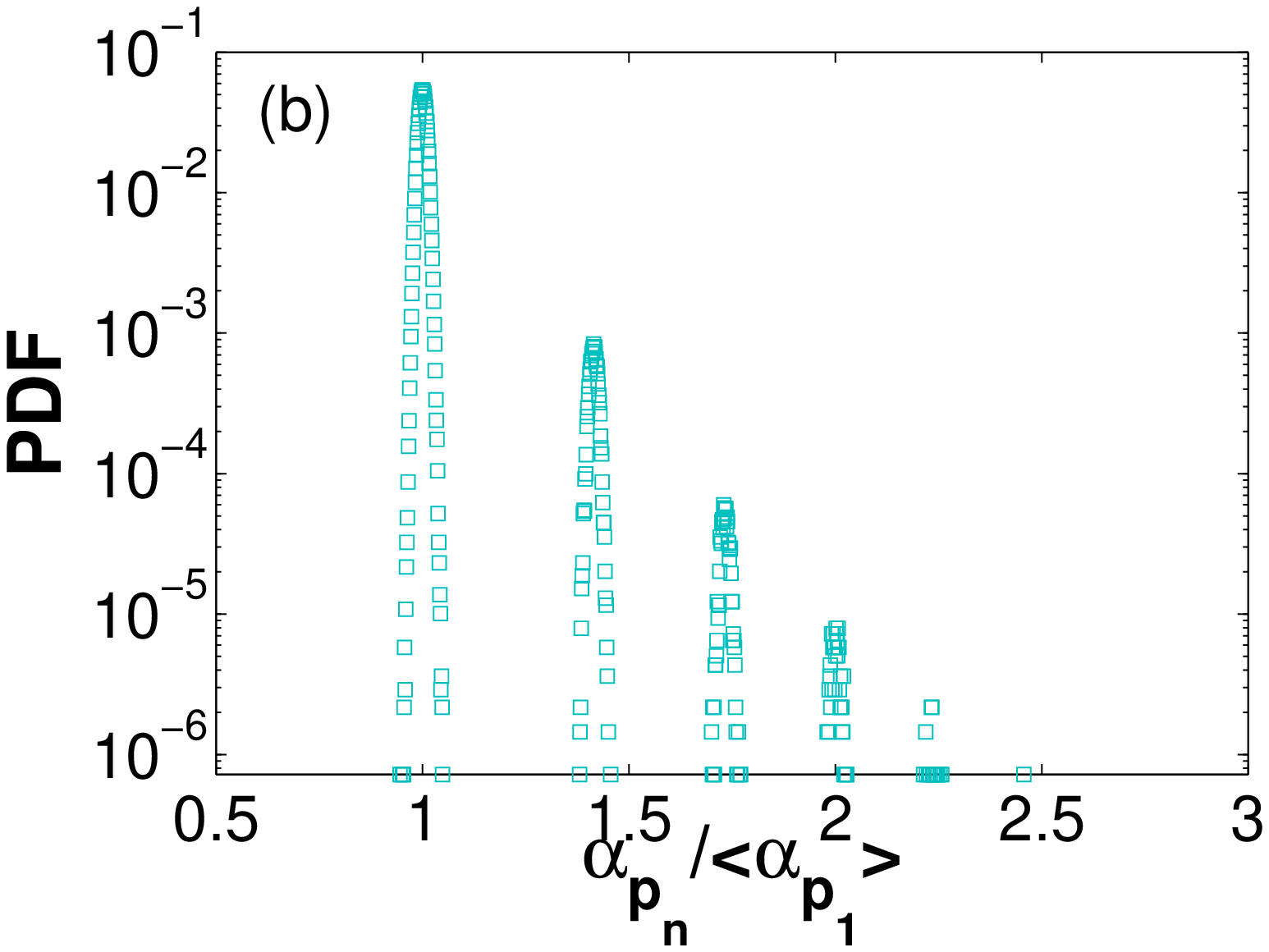}
  \includegraphics[width=0.235\textwidth]{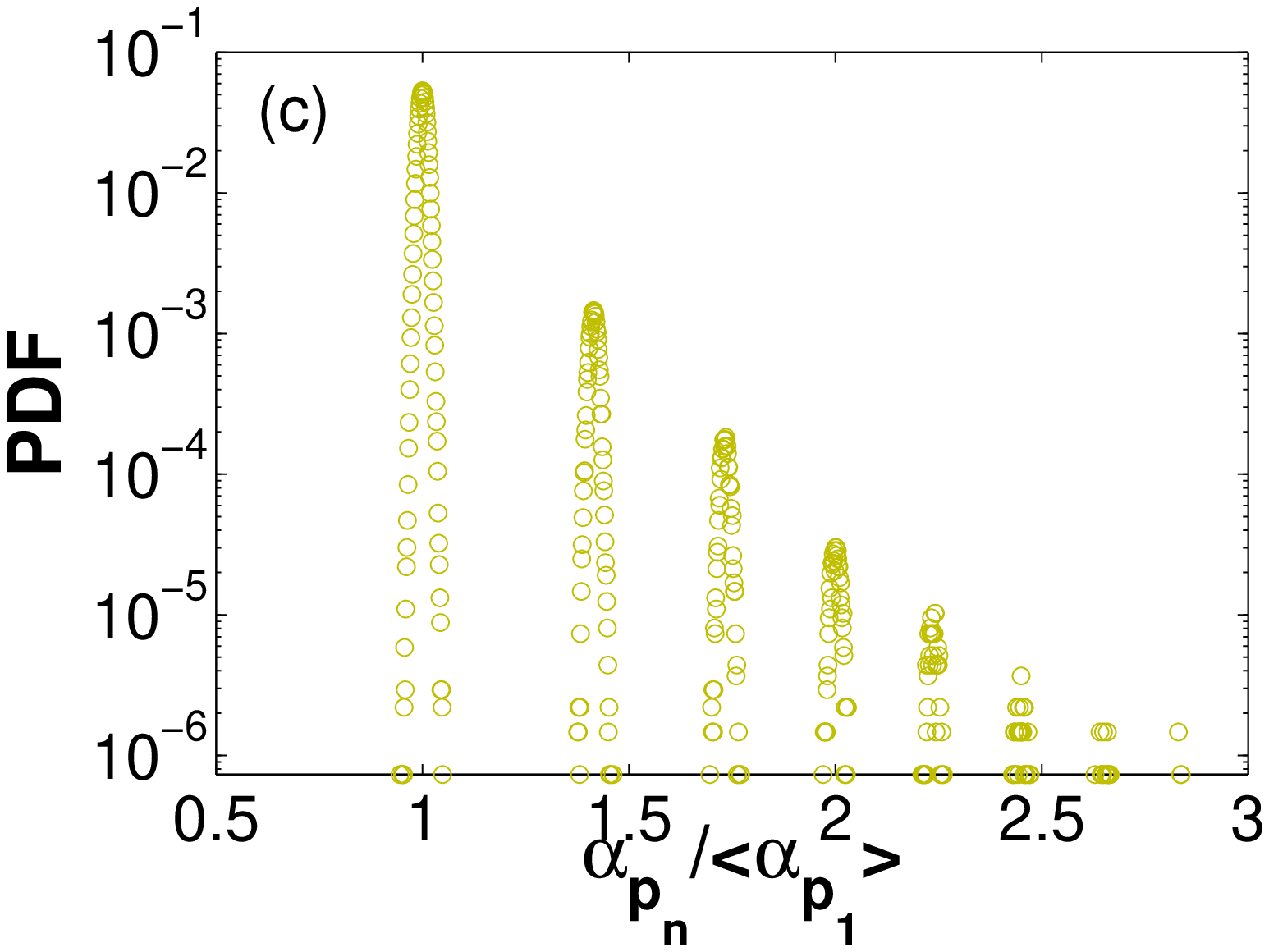}
  \includegraphics[width=0.235\textwidth]{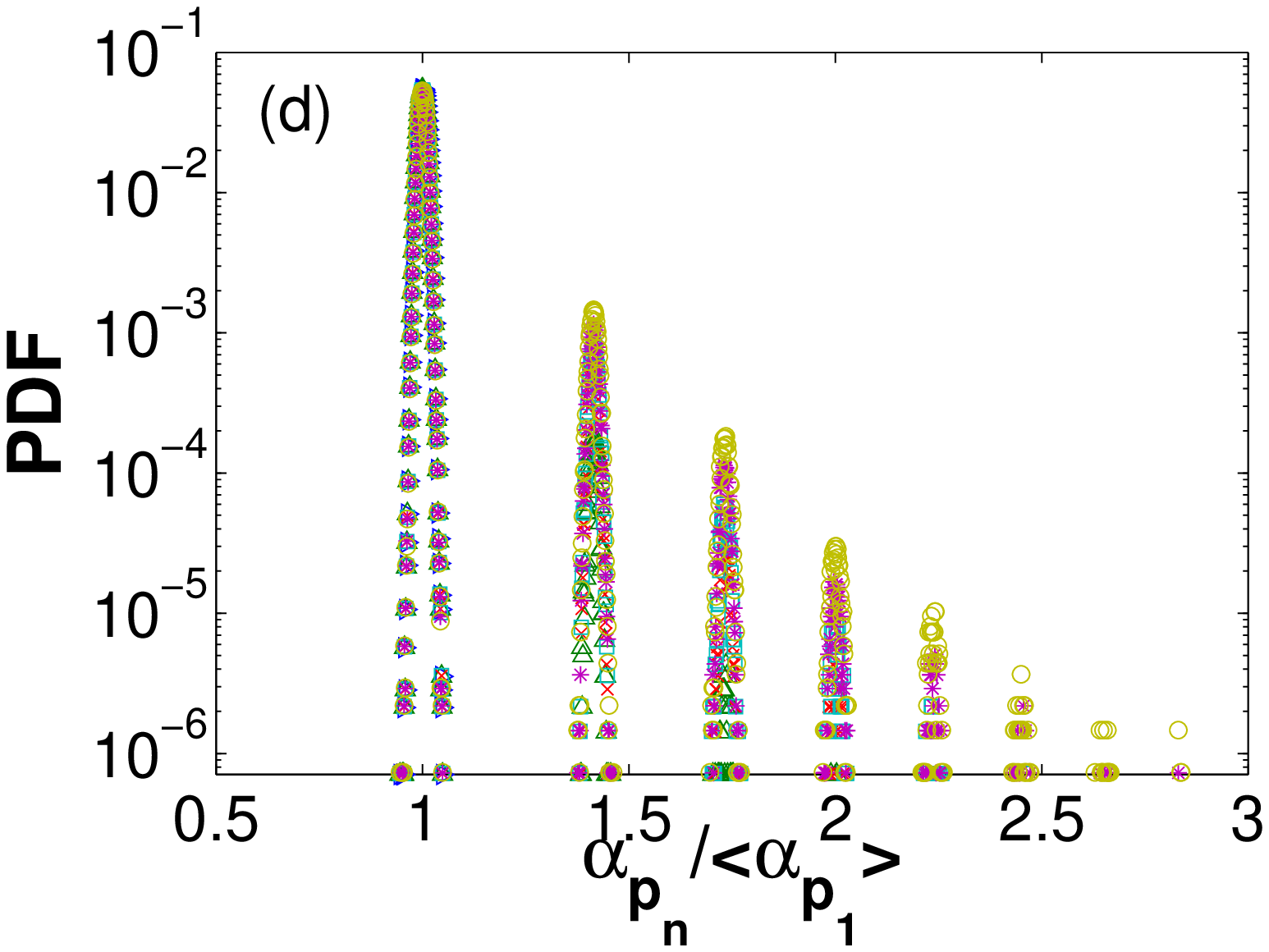}
  \caption{(Color online) PDFs of the particle radius $a_p$ normalised by the initial
    average particle radius $\avg{a_p}$ at different times; in plot (a) time
    $3\tau_L$, (b) time $11.6 \tau_L$, (c)
    time $22\tau_L$. In plot (d), six
    PDFs are plotted together at six different times: blue $\vartriangleright 0.0 \tau_L$, green $\vartriangle 3.0 \tau_L$, red $\times \; 8.5 \tau_L$, cyan $\square \; 11.6 \tau_L$, magenta $* \; 17.0 \tau_L$, yellow $\circ \; 22 \tau_L$.}
  \label{fig:dropgrowth}
\end{figure}

Figure \ref{fig:dropgrowth} shows PDFs of particle radii at different
times. At $t = 0$, this PDF is very sharply peaked around $0.0135\eta$. By
$t = 3\tau_L$ a second peak has appeared at a value about 1.4 times
larger than $0.0135\eta$, and as time progresses more peaks appear at
progressively higher values of the particle radius. By the end of the
simulation, i.e. when $t/\tau_{L} = 22$, the largest particle has a
radius nearly three times larger than the initial particle radii. This
result of our simulations is noteworthy because $t/\tau_{L} = 22$
corresponds to about less than a minute and the factor 3 is not too
far from the ratio of $50\mu$m to $15\mu$m.


In conclusion, our calculations suggest that turbulence in air which
is carrying water droplets so small that their average Stoker number
is of order $10^{-2}$ and therefore too small to significantly cluster
can nevertheless initiate a process of droplet growth via collisions
and coalescence which is fast enough for droplets to grow from
$15\mu$m to $50\mu$m within about a minute. The reason behind this
fast growth is in the wide spread of local Stokes numbers in a
turbulent flow. A few local flow events exist in the turbulent field
where the local Stokes number is much higher than the average Stokes
number and high enough to cause a few pairs of droplets to collide and
coalesce quite quickly even if small-scale intermittency and
differential gravitational settling are not taken into
account. However, this initiation mechanism may not be effective if
the volume fraction is too low. When we took the simulated volume
fraction $\phi_V$ to be smaller than $10^{-3}$, i.e.  $\phi_V
\approx 2 \cdot 10^{-4}$ and $\phi_V \approx 3 \cdot 10^{-5}$, the
first ten collisions returned by our simulations were at significantly
later times and not all local Stokes numbers in these collisions were
significantly larger than average (6 and 3 pairs, respectively, in the
$\phi_V \approx 2\cdot 10^{-4}$ and $\phi_V \approx 3\cdot 10^{-5}$
cases). Nevertheless, we stress that we have found a
turbulence-generated droplet growth phenomenon which takes droplets
from $15\mu$m to $50\mu$m within a time which, if multiplied by our
estimated a priori frequency of collisions, is comparable to the time
required for droplets to grow from $15\mu$m to $50\mu$m in warm
clouds.


Our results support Kostinski's \& Shaw's \cite{kostinskishaw05} 
suggestion that powerful rare events can cause
initiation of fast droplet growth by turbulence. However, our results
also suggest that this phenomenon may not require small-scale
intermittency and/or differential gravitational settling if collision
efficiency is assumed perfect. Nevertheless, one can surely expect
selectively enhanced settling velocities \cite{ghoshetal05} and/or
small-scale intermittency to increase the number and/or power of
events which can accelerate average growth rate and thereby perhaps
outweigh the adverse effect of collision efficiencies which are
typically one order of magnitude smaller than assumed in this work.




%
The authors are grateful to Ryo Onishi, Wojciech Grabowski and
Bernhard Mehlig for reading the first version of this manuscript and
making very helpful suggestions and comments.
%
\bibliography{refs}
\end{document}